\documentclass[aps,floatfix,showpacs,superscriptaddress,preprint]{revtex4}
\usepackage{latexsym,graphicx,epsfig,psfrag}

\usepackage{epstopdf}
\usepackage{amssymb,amsmath,amsxtra,amsfonts}
\usepackage{bm}

\usepackage{longtable}
\usepackage{multirow}
\usepackage{booktabs}
\usepackage{array}
\usepackage{wrapfig}
\usepackage{color}
\usepackage{titlesec}

\titleformat{\section}{\large\bfseries}{\thesection}{1em}{}

\newcommand{\bea}{\begin{eqnarray}}
\newcommand{\ena}{\end{eqnarray}}
\newcommand{\be}{\begin{equation}}
\newcommand{\en}{\end{equation}}
\newcommand{\nn}{\nonumber\\}
\newcommand{\ed}{\end{document}} 
\newcommand{\Tr}{\mbox{\rm{tr}}}

\newcommand{\la}{\langle}
\newcommand{\ra}{\rangle}

\begin{document}

\title{$B^{*}_c$ meson parameters and radiative decay width within the covariant confined quark model} 

\author{Aidos Issadykov}
\email{issadykov.a@gmail.com}

\author{Sayabek K. Sakhiyev}

\affiliation{The Institute of Nuclear Physics, \\
Ministry of Energy of the Republic of Kazakhstan, 050032 Almaty,  {\it KAZAKHSTAN}}

\begin{abstract}
In this work we tried to predict the parameters of $B^{*}_c$ meson. Simple assumptions gave us following parametres $m_{B_{c}^{*}}=6329\pm 10$ MeV and $f_{B_{c}^{*}}= 535.5\pm57.8$ MeV (for $\Lambda_{B_{c}^{*}}=2.26\pm 0.14$ GeV in covariant confined quark model). We calculated widths of radiative decays of  $B^*_{q}$ mesons, where $q=u/d,s,c$ and compared them with other theoretical works. It was shown that the width of the $B_{c}^{*}$ meson very sensitive to the mass $m_{B_{c}^{*}}$ as expected and less to the size parameter $\Lambda_{B_{c}^{*}}$.
\end{abstract}


\maketitle
\section{Introduction}

The decay mode $ B_{c} \to J/\psi \ell \nu $ of $B_c$ meson have about 2 standard deviations disagreement between experimental data and theoretical predictions~\cite{Aaij:2017tyk}. Meanwhile, its vector partner $ B_c ^*$ is still not found. It is expected that the mass difference is not large to decay strongly to $ B_c $ meson and light meson. Thus,
 $ B_c ^*$ mesons cannot decay strongly but can decay only weakly and electromagnetically. As a result, the partial widths of electromagnetic decay channels, especially single-photon decay channels, are dominant.
Since the $B^{*}_c$ meson was not observed yet, there are some theoretical predictions of it's mass and leptonic decay constants in the relativistic quark model\cite{Ebert:2002pp}, Lattice QCD\cite{Dowdall:2012ab,Colquhoun:2015oha}, QCD Sum Rules\cite{Wang:2012kw} and Nonrelativistic renormalization group\cite{Penin:2004xi}.
Properties of $B^{*}_c$ meson in the relativistic quark model\cite{Ebert:2002pp} as follows:
\bea
m_{B_{c}^{*}}=6332 \quad ~\text{MeV},  \qquad f_{B_{c}^{*}}= 503 \quad ~\text{MeV}.
\ena
Mass and leptonic decay constant of $B^{*}_c$ meson in Lattice QCD\cite{Dowdall:2012ab,Colquhoun:2015oha} looks like:
\bea
m_{B_{c}^{*}}=6332 \pm 9 \quad ~\text{MeV},  \qquad f_{B_{c}^{*}}= 422\pm 13 \quad ~\text{MeV}.
\ena

Mass and leptonic decay constant of $B^{*}_c$ meson from QCD Sum Rules\cite{Wang:2012kw}:
\bea
m_{B_{c}^{*}}=6337 \quad ~\text{MeV},  \qquad f_{B_{c}^{*}}= 384 \quad ~\text{MeV}.
\ena

The Nonrelativistic renormalization group~\cite{Penin:2004xi} gave their prediction on mass differences of $B^{*}_c$ and $B_c$ mesons
\be
\Delta m_{({B_{c}^{*}-B_{c}})}=50\pm17 ^{+15}_{-12} \quad ~\text{MeV}.
\en

Radiative decay of $B_c^*$ meson was calculated in \cite{Chang:2020xvu,Simonis:2018rld,Jena:2002is,Priyadarsini:2016tiu,Patnaik:2017cbl,Ebert:2002xz,Ebert:2002pp,Lahde:1999ih,Lahde:2002wj,Choi:2007se,Choi:2009ai,Eichten:1994gt,Kiselev:1994rc,Fulcher:1998ka,Nobes:2000pm,Monteiro:2016rzi,AbdElHady:2005bv} and have partial widths less than 1 keV which makes the branching ratios
of their weak decay modes may be within the detection ability of current experiments.
There are several works dedicated to investigate the semileptonic decays of $B_c^*$ ~\cite{Wang:2012hu,Dai:2018vzz,Wang:2018ryc,Chang:2020xvu}.
The purpose of this paper is to extend our model and predict a model parameters of unobserved $B_c^*$.
We studied $b \to c$, $b \to s$ and $b \to d(u)$ transitions in the framework of covariant confined quark model(CCQM) in our previous works\cite{Soni:2021fky,Soni:2020bvu,Issadykov:2018myx,Dubnicka:2016nyy,Issadykov:2015iba}.

\section{Model}
\label{sec:model}

The covariant confined quark model 
\cite{Efimov:1988yd,Efimov:1993ei,Branz:2009cd}
is an effective quantum field approach to hadronic interactions based on 
an interaction Lagrangian of hadrons interacting with their constituent quarks.

The effective Lagrangian describing
the transition of a meson $M(q_1\bar q_2)$ to its constituent
quarks $q_1$ and $\bar q_2 $ 
\bea
{\mathcal L}_{\rm int}(x) &=& g_M M(x)\cdot J_M(x) + {\rm h.c.},
\nn
J_M(x) &=& \int\!\! dx_1 \!\!\int\!\!
dx_2 F_M (x,x_1,x_2)\bar q_2(x_2)\Gamma_M q_1(x_1) 
\label{eq:strong-int-Lag}
\ena
with $\Gamma_M$ a Dirac matrix which projects onto the spin quantum 
number of the meson field $M(x)$. The vertex function $F_M$  characterizes 
the finite size of the meson. Translational invariance requires the 
function $F_M$ to fulfill the identity $F_M(x+a,x_1+a,x_2+a)=F_M(x,x_1,x_2)$ for
any four-vector $a$. A specific form for the  vertex function is adopted
\be
F_M(x,x_1,x_2)=\delta(x - w_1 x_1 - w_2 x_2) \Phi_M((x_1-x_2)^2),
\label{eq:vertex}
\en
where $\Phi_M$ is the correlation function of the two constituent quarks 
with masses $m_{q_1}$ and $m_{q_2}$. The ratios of the quark masses
$w_i$ are defined as 
\be
w_{q_1}=\frac{m_{q_1}}{m_{q_1}+m_{q_2}}, \quad
w_{q_2}=\frac{m_{q_2}}{m_{q_1}+m_{q_2}}, \quad w_1+w_2=1.
\label{eq:w_i}
\en

A simple Gaussian form of the vertex function $\bar \Phi_M(-\,k^2)$ is selected
\be
\bar \Phi_M(-\,k^2) 
= \exp\left(k^2/\Lambda_M^2\right)
\label{eq:Gauss}
\en
with the parameter $\Lambda_M$ linked to the size of the meson. The minus sign 
in the argument is chosen to indicate that we are working in the Minkowski 
space. Since $k^2$ turns into $-\,k_E^2$ in the Euclidean space, 
the form (\ref{eq:Gauss}) has the appropriate fall-off behavior in 
the Euclidean region. Any choice for  $\Phi_M$ is appropriate
as long as it falls off sufficiently fast in the ultraviolet region of
the Euclidean space to render the corresponding Feynman diagrams ultraviolet 
finite. We choose a Gaussian form for calculational convenience.

The coupling constant $g_M$ in Eq.~(\ref{eq:strong-int-Lag}) is determined
by the so-called {\it compositeness condition}. 
The compositeness condition requires that the renormalization constant $Z_B$ 
of the elementary meson field $B(x)$ is set to zero, i.e.
\be
\label{eq:Z=0}
Z_B=1-\widetilde\Pi^\prime_B(p^2)=0, \qquad (p^2=m^2_B)
\en
where $\Pi^\prime_B(p^2)$ is the derivative of the mass function.

$S$-matrix elements are described by the quark-loop diagrams
which are the convolution of the vertex functions and quark
propagators. In the evaluation of the quark-loop diagrams we
use the local Dirac propagator
\be
S_q(k) = \frac{1}{ m_q-\not\! k -i\epsilon } = 
\frac{m_q + \not\! k}{m^2_q - k^2  -i\epsilon }
\label{eq:prop}
\en 
with an effective constituent quark mass $m_q$.

The meson functions in the case of the pseudoscalar and vector meson
are written as
\bea
\widetilde\Pi_P(p^2) &=& N_c g_P^2
\int\frac{d^4k}{(2\pi)^4i} \widetilde\Phi^2_P(-k^2)
\Tr\Big(\gamma^5 S_1(k+w_1 p)\gamma^5 S_2(k-w_2 p)\Big),
\label{eq:massP}\\[2ex]
\widetilde\Pi^{\mu\nu}_V(p^2) &=&N_c g_V^2 
\int\frac{d^4k}{(2\pi)^4i} \widetilde\Phi^2_V(-k^2)
\Tr\Big(\gamma^\mu S_1(k+w_1 p)\gamma^\nu S_2(k-w_2 p)\Big)
\nn
&=& g^{\mu\nu} \widetilde\Pi_V(p^2) + p^\mu p^\nu \widetilde\Pi^\parallel_V(p^2).
\label{eq:massV}
\ena
Here  $N_c=3$ is the number of colors.
Since the vector meson is
on its mass-shell  $\epsilon_V\cdot p=0$ we need to keep
the part $ \widetilde\Pi_V(p^2)$. Substituting the derivative of the mass
functions into Eq.~(\ref{eq:Z=0}) one can determine the coupling
constant $g_B$ as a function of other model parameters.
The loop integrations in Eqs.~(\ref{eq:massP}) and ~(\ref{eq:massV})
proceed by using the Fock-Schwinger representation
of quark propagators
\be
S_q (k+w p) = \frac{1}{ m_q-\not\! k- w \not\! p } 
= (m_q + \not\! k + w \not\! p)\int\limits_0^\infty \!\!d\alpha\, 
e^{-\alpha [m_q^2-(k+w p)^2]}.
\label{eq:Fock}
\en
In the obtained integrals over the Fock-Schwinger parameters 
$0\le \alpha_i<\infty$
we introduce an additional integration over the proper time
which converts the set of 
Fock-Schwinger parameters into a simplex. In general case one has
\be
\prod\limits_{i=1}^n\int\limits_0^{\infty} 
\!\! d\alpha_i f(\alpha_1,\ldots,\alpha_n)
=\int\limits_0^{\infty} \!\! dtt^{n-1}
\prod\limits_{i=1}^n \int\!\!d\alpha_i 
\delta\left(1-\sum\limits_{i=1}^n\alpha_i\right)
  f(t\alpha_1,\ldots,t\alpha_n).
\label{eq:simplex}  
\en
Finally, we cut the integration over the proper time
at the upper limit by introducing an infrared cutoff $\lambda$.
One has
\be
\int\limits_0^\infty dt (\ldots) \to \int\limits_0^{1/\lambda^2} dt (\ldots).
\label{eq:conf}
\en
This procedure allows us to remove all possible thresholds present
in the initial quark diagram. Thus the infrared cutoff parameter 
$\lambda$ effectively guarantees the confinement of quarks within hadrons. 
This method is quite general and can be used for diagrams with an arbitrary 
number of loops and propagators. 
In the CCQM the infrared cutoff parameter $\lambda$ is taken to be universal 
for all physical processes.

The model parameters are determined by fitting calculated quantities 
of basic processes to available experimental data or 
lattice simulations (for details, see Ref.~\cite{Branz:2009cd}).

\section{Matrix elements and one-photon radiative decay width}
\label{sec:decaywidth}

\begin{figure}
\begin{center}
\includegraphics[scale=0.2]{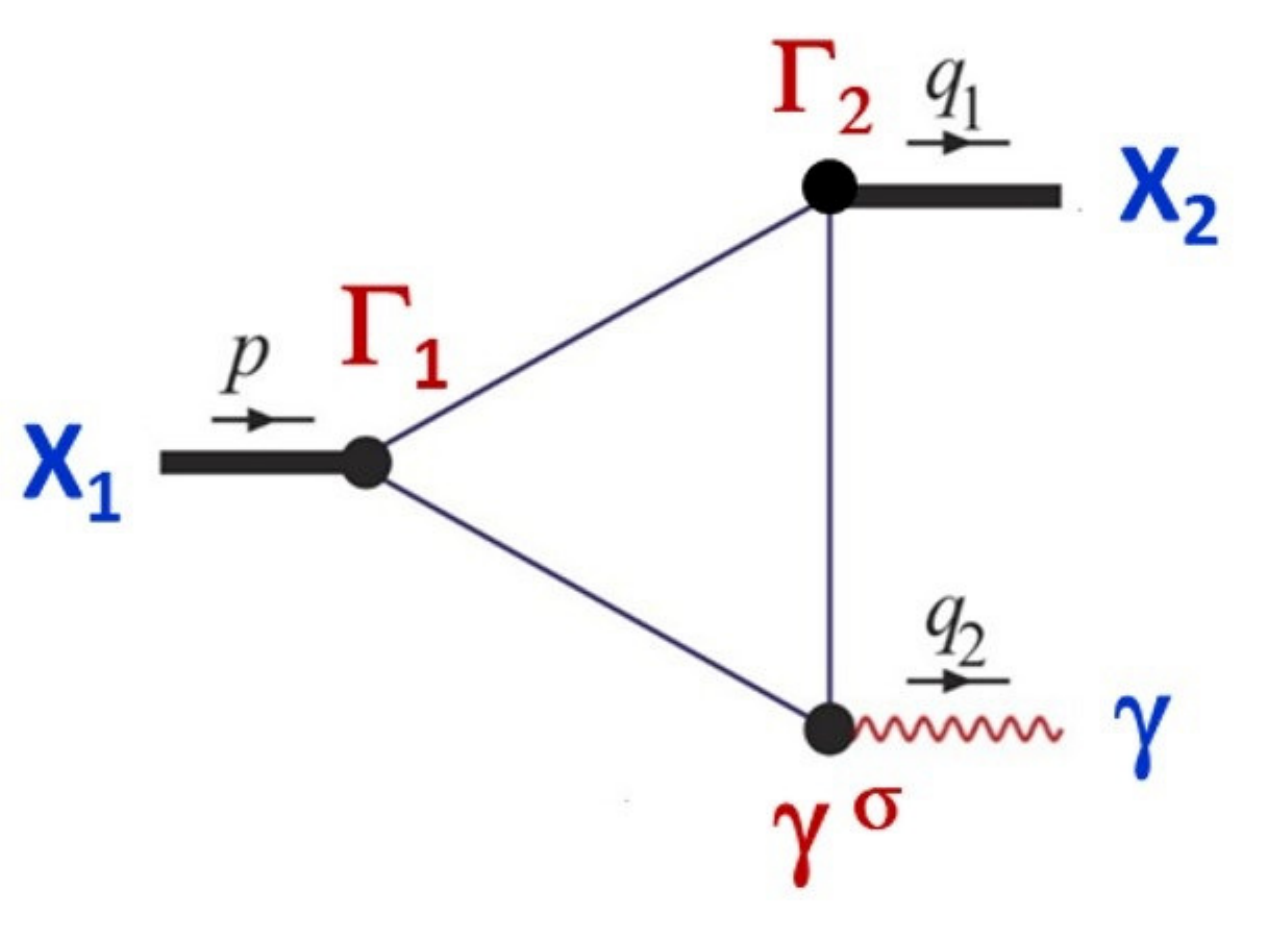}
\hspace*{2.5mm}
\includegraphics[scale=0.2]{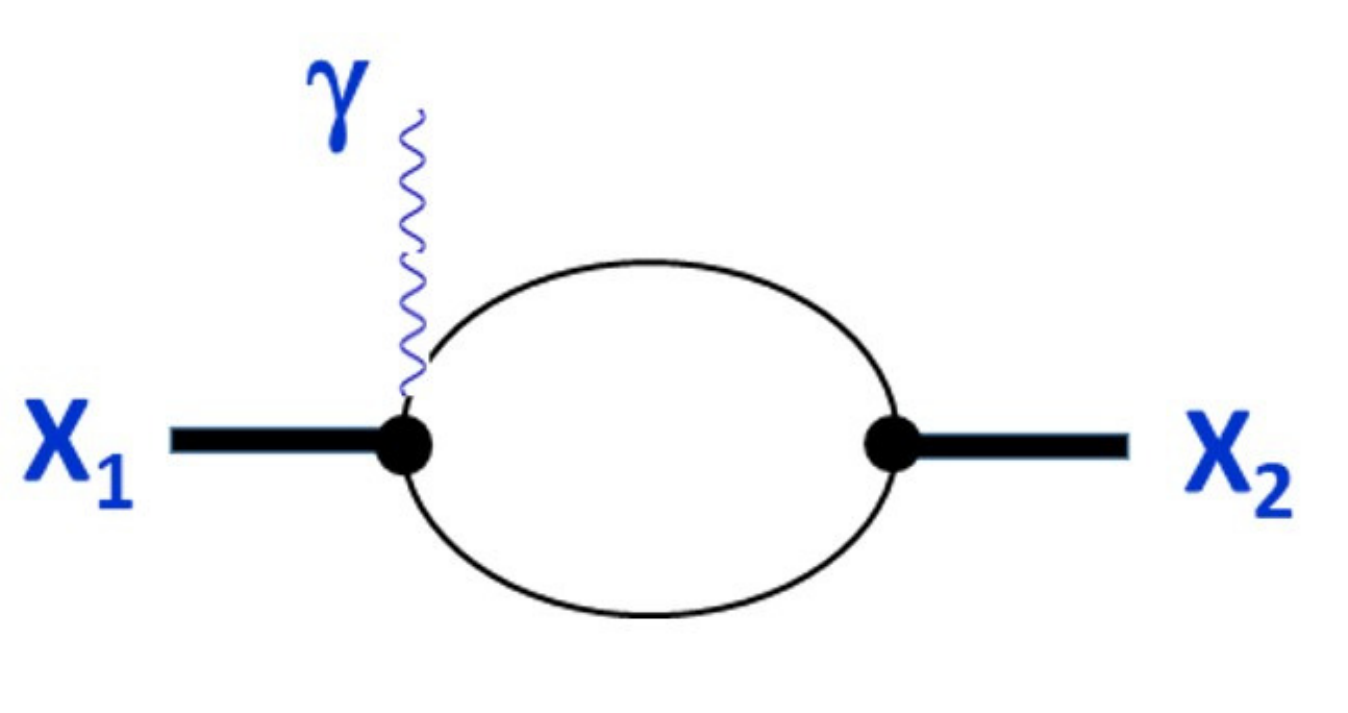}
\hspace*{2.55mm}
\includegraphics[scale=0.2]{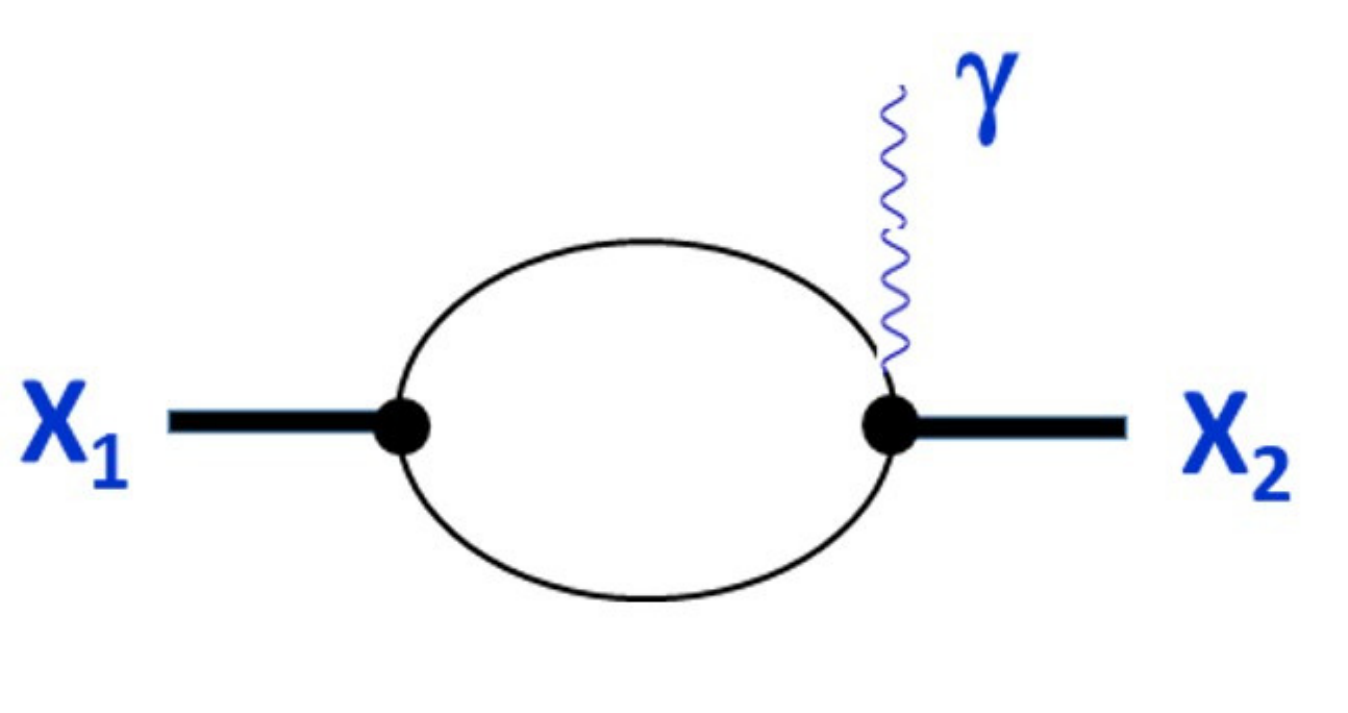}
\end{center}
\caption{
Feynman diagrams contributing in
leading order to the dominant one-photon radiative transitions
$X_{1}(p)\to\gamma(q_2)+X_{2}(q_1)$ \cite{Ganbold:2021nvj}.
}
\label{fig1}
\end{figure}

The free Lagrangian of quarks 
is gauged in the standard manner by using minimal substitution which
gives
\be
\mathcal{L}^{\rm em}_{\rm int}(x) = e\,A_\mu(x)\,J^\mu_{\rm em}(x),\qquad
J^\mu_{\rm em}(x)= e_b\,\bar{b}(x)\gamma^\mu b(x)
+e_q\,\bar{q}(x)\gamma^\mu q(x)
\label{eq:em-lag}
\en
where $e_b$ and $e_q$ are the quark charges in units of the positron charge.
The radiative decays of a vector mesons into a pseudoscalar meson
and photon $X_1\to X_2\gamma$ are described by the Feynman diagrams shown
in Fig.~\ref{fig1}.

The invariant matrix element for the one-photon radiative transition
$X_{1} \to \gamma X_{2}$ reads
\begin{eqnarray}
{\cal M}_{{X_1} \to \gamma {X_2}} (p;p',q) = e g_{X_1} g_{X_2} \epsilon^V_\nu(p)\epsilon^\gamma_\mu(q)
\int\!\! dx\!\!  \int\!\!  dy\!\!  \int\!\!  dz\,
e^{ -ipx + ip^{\prime}y + iqz}
\la\,  T \{ \bar{J}_{X_1}^\nu (x)  J^\mu_{\rm em} (z) J_{X_2} (y)  \} \ra_0.,
\label{matrix}
\end{eqnarray}

One has to note that there is an additional piece in the Lagrangian
 related to the gauging nonlocal interactions of hadrons with
 their constituents~\cite{Branz:2009cd}. This piece gives the additional
 contributions to the electromagnetic processes. However, they are identically
 zero for the process  $X_1\to X_2\gamma$ due to its anomalous nature. 

Using the Fourier transforms of the quark currents, we come to the final result
\bea
{\cal M}_{{X_1} \to \gamma {X_2}} (p;p',q) &=& (2\pi)^4 i\, \delta(p-p'-q) M(p,p'),
\nn
M(p, p') &=& (-3i) e g_{X_1} g_{X_2} \epsilon^V_\nu(p)\epsilon^\gamma_\mu(q)\,
\left( e_b M^{\mu\nu}_b + e_q M^{\mu\nu}_q\right)
\nn
M^{\mu\nu}_b &=&
\int\!\!\frac{dk}{(2\pi)^4 i} \widetilde\Phi_{X_1}(-\ell_1^2)
                              \widetilde\Phi_{X_2}(-\ell_2^2)
\Tr\left[S_q(k)\gamma^\nu S_b(k-p) \gamma^\mu S_b(k-p')\gamma^5 \right]
\nn
M^{\mu\nu}_q&=& \int\!\!\frac{dk}{(2\pi)^4 i}\widetilde\Phi_{X_1}(-\ell_3^2)
                                  \widetilde\Phi_{X_2}(-\ell_4^2) 
\Tr\left[S_q(k+p')\gamma^\mu S_q(k+p)\gamma^\nu S_b(k)\gamma^5 \right]
\ena
where $\ell_1=k-w_2\, p$, $\ell_2=k-w_2\, p'$ and
$\ell_3=k+w_1\, p$, $\ell_2=k+w_1\, p'$. The ratios of quark masses
are defined by Eq.~(\ref{eq:w_i}). Now one has
$m_{q_1}=m_b$ and $m_{q_2}=m_q$ with $q=u,d,s$.
By using the technique of calculations and taking into account the transversality conditions
$\epsilon^\gamma_\mu(q)q^\mu=0$ and $ \epsilon^V_\nu(p)p^\nu=0$
one can arrives at the standard form of matrix element
\be
M(p,p') = e\, g_{X_1 X_2 \gamma}\,\varepsilon^{p q \mu \nu} \epsilon^\gamma_\mu(q)
\epsilon^V_\nu(p),
\en
where $ g_{X_1 X_2 \gamma} =  e_b I_b(m^2_{X_1},m^2_{X_2}) + e_q I_q(m^2_{X_1},m^2_{X_2}) $
is radiative decay constant. The quantities $I_{b,q}$ are defined by
the two-fold integrals which are calculated numerically.
The electromagnetic decay width is written as
\be
\Gamma(X_{1}\to X_{2} + \gamma) =
\frac{\alpha}{24} m_{X_1}^3\left(1-\frac{m_{X_2}^2}{m_{X_1}^2}\right)^3 g_{X_1 X_2\gamma}^2\,.
\en
where $\alpha=e^2/4\pi =1/137.036$ is the fine-structure constant. 

\section{Numerical results}


The obvious model parameters include constituent quark masses and meson size parameters that are fixed by fitting with the basic processes such as leptonic decay widths with the experimental data or lattice simulations and the differences are considered to be the absolute uncertainty in the respective parameter.
These parameters are determined by
minimizing the functional
$\chi^2 = \sum\limits_i\frac{(y_i^{\rm expt}-y_i^{\rm theor})^2}{\sigma^2_i}$
where $\sigma_i$ is the experimental  uncertainty.
If $\sigma$ is too small then we take its value of 10$\%$.
Besides, we have observed that the errors of the fitted parameters 
are of the order of  10$\%$.
Thus, the theoretical error of the CCQM is estimated to be of the order
of 10$\%$ at the level of matrix elements and the order
of 15$-$20$\%$ at the level of widths.
For present computations, we use the model parameters obtained using the updated least square fit method performed in the Ref. \cite{Ivanov:2015tru,Ganbold:2014pua,Dubnicka:2016nyy}.

\begin{table}[ht]
\begin{center}
\def\arraystretch{1.2}
\caption{Input values for some basic electromagnetic decay widths and our
least-squares  fit values (in keV).}
\label{tab:em-widths}
\vspace*{0.2cm}
\begin{tabular}{lll}
\hline\hline
Process & Fit Values & Data~\cite{ParticleDataGroup:2020ssz}   \\
\hline
$\rho^{\pm}\to\pi^{\pm}\gamma$    & 75.7$\pm$ 15.1     &  67 $\pm$ 7.5   \\
$\omega\to\pi^0\gamma$          & 679$\pm$ 135.8      &  713 $\pm$ 26      \\
$K^{\ast \pm}\to K^\pm\gamma$      & 55.8$\pm$ 11.2     &  46.8 $\pm$ 4.7  \\
$K^{\ast 0}\to K^0\gamma$         & 132$\pm$ 26.4      &  116 $\pm$ 10      \\
$D^{\ast \pm}\to D^\pm\gamma$      & 0.75$\pm$ 0.15     &  1.33 $\pm$ 0.37   \\
$J/\psi \to \eta_c \gamma $      & 1.77$\pm$ 0.35     &  1.58 $\pm$ 0.37       \\
\hline
\end{tabular}
\end{center}
\end{table}
The results of the least-squares fit used in the present study can be found in Table~\ref{tab:em-widths}. The agreement between the fit and experimental data is quite satisfactory.
The result for $J/\psi \to \eta_c \gamma $ agrees with the one given in \cite{Ganbold:2021nvj}(please look Table II there).

We think that there are strong relation between pseudoscalar $B_{q}$ and vector $B^{*}_{q}$ mesons.
In Table~\ref{eq:lept} given the leptonic decay constants and masses of $B_{q}^{(*)}$ mesons from PDG~\cite{ParticleDataGroup:2020ssz} and corresponding fitted size parameters from previous works in CCQM~\cite{Issadykov:2015iba,Dubnicka:2016nyy,Dubnicka:2017job,Issadykov:2017wlb,Issadykov:2018myx}.

The leptonic decay constants in CCQM are defined by Eq.10 in ~\cite{Issadykov:2017wlb}.

\bgroup 
\begin{table}[htp]
\caption{\label{eq:lept}
The values of the leptonic decay constants and meson masses(in MeV)  except the $B^{*}_c$ meson parameters from PDG~\cite{ParticleDataGroup:2020ssz} and corresponding our model parameter $\Lambda$(in GeV)from our previous works~\cite{Issadykov:2015iba,Dubnicka:2016nyy,Dubnicka:2017job,Issadykov:2017wlb,Issadykov:2018myx}. }
\vspace*{2mm}  
\centering
\def\arraystretch{1.5}
 \begin{tabular}{|c|c|c|c|c|c|c|}
\hline
  & $B_{c}$           &$ B^{*}_s$               &$B_s$   &$B^{*0}$   &$B^{0}$        &$B^{+}$  \\
\hline
$m$         &$6274.47\pm 0.32$   &$5415.4 ^{+1.8} _{-1.5}$    &$5366.88\pm 0.14$ &$5324.70\pm 0.21$     &$5279.65\pm 0.12$         &$5279.34\pm 0.12$\\
$f$               &489         &229      &238.7        & 196        &193           &193 \\
$\Lambda$         &2.73         &1.79      &2.05        & 1.80        &1.96           &1.96 \\
\hline
\end{tabular}
\end{table}
\egroup

From Table~\ref{eq:lept} one can find next mass differences between pseudoscalar and vector mesons
\begin{eqnarray}
\Delta m_{({B_{s}^{*}-B_{s}})}=49 \quad ~\text{MeV}, \\
\nn
\Delta m_{({B^{*0}-B^{0}})}=45 \quad ~\text{MeV},
\end{eqnarray}

so that the mass for $B_{c}^{*}$ meson assumed as:
\bea
\Delta m_{({B_{c}^{*}-B_{c}})}=55\pm 10 \quad ~\text{MeV,}
\quad  ~\text{then} \quad  m_{B_{c}^{*}}=6329\pm 10 \quad ~\text{MeV,}
\ena
which is within the predictions of other models\cite{Ebert:2002pp,Dowdall:2012ab,Colquhoun:2015oha,Wang:2012kw,Penin:2004xi}.

The ratio between size parameters of $B_{q}^{(*)}$ mesons from our previous works~\cite{Issadykov:2015iba,Dubnicka:2016nyy,Dubnicka:2017job,Issadykov:2017wlb,Issadykov:2018myx} as follows

\begin{eqnarray}
\Delta \Lambda_{({B_{s}^{*}/B_{s}})}=0.876, \\
\nn
\Delta \Lambda_{({B^{*0}/B^{0}})}=0.921,
\end{eqnarray}

so that the size parameter $\Lambda_{B_{c}^{*}}$ assumed as:

\bea
\Delta \Lambda_{({B_{c}^{*}/B_{c}})}=0.83\pm 0.05,
\quad  ~\text{then} \quad  \Lambda_{B_{c}^{*}}=2.26\pm 0.14 \quad ~\text{GeV.}
\ena

Taking into account these two parameters we calculated the width of radiative decay $\Gamma (B^{*+}_c \to B^{+}_c \gamma)$ and  $f_{B^{*}_c}$ leptonic decay constant in In Table~\ref{tab:Bvc}.
We calculated the widths of radiative decay in dependence from mass(6319$-$6339 MeV) and $\Lambda$(2.12$-$2.40 GeV) parameters of $B^*_{c}$ meson.
\begin{table}[htp] 
\caption{\label{tab:Bvc}
The widths of radiative decay of  $B^*_{c}$ meson in dependence from mass and $\Lambda$ parameters. }
\begin{center}
\begin{tabular}{|c|c|c|}
\hline

\hline
$m_{B_{c}^{*}}=6319$ $~\text{MeV}$ &$\Gamma (B^{*+}_c \to B^{+}_c \gamma), ~\text{(keV)}$ &$f_{B_{c}^{*}},~\text{(MeV)}$  \\
\hline
$\Lambda=2.12$   & 0.023  & 481\\
$\Lambda=2.19$   & 0.024  & 508.5\\
$\Lambda=2.26$   & 0.025  & 536.4\\
$\Lambda=2.33$   & 0.026  & 564.6\\
$\Lambda=2.40$   & 0.027  & 593.3\\
\hline
\hline
$m_{B_{c}^{*}}=6324$ $~\text{MeV}$ &$\Gamma (B^{*+}_c \to B^{+}_c \gamma), ~\text{(keV)}$ &$f_{B_{c}^{*}},~\text{(MeV)}$  \\
\hline
$\Lambda=2.12$   & 0.032  & 479.9\\
$\Lambda=2.19$   & 0.033  & 507.3\\
$\Lambda=2.26$   & 0.034  & 535\\
$\Lambda=2.33$   & 0.035  & 563.1\\
$\Lambda=2.40$   & 0.036  & 591.6\\
\hline
\hline
$m_{B_{c}^{*}}=6329$ $~\text{MeV}$ &$\Gamma (B^{*+}_c \to B^{+}_c \gamma),~\text{(keV)}$ &$f_{B_{c}^{*}},~\text{(MeV)}$  \\
\hline
$\Lambda=2.12$   & 0.042  & 478.8\\
$\Lambda=2.19$   & 0.044  & 506\\
$\Lambda=2.26$   & 0.045  & 533.6\\
$\Lambda=2.33$   & 0.047  & 561.6\\
$\Lambda=2.40$   & 0.048  & 589.9\\
\hline
\hline
$m_{B_{c}^{*}}=6339$ $~\text{MeV}$ &$\Gamma (B^{*+}_c \to B^{+}_c \gamma),~\text{(keV)}$ &$f_{B_{c}^{*}},~\text{(MeV)}$  \\
\hline
$\Lambda=2.12$   & 0.069  & 476.5\\
$\Lambda=2.19$   & 0.072  & 503.5\\
$\Lambda=2.26$   & 0.074  & 530.8\\
$\Lambda=2.33$   & 0.077  & 558.5\\
$\Lambda=2.40$   & 0.079  & 586.5\\
\hline
\end{tabular}
\end{center}
\end{table}

\begin{figure}
\begin{center}
\includegraphics[scale=1.]{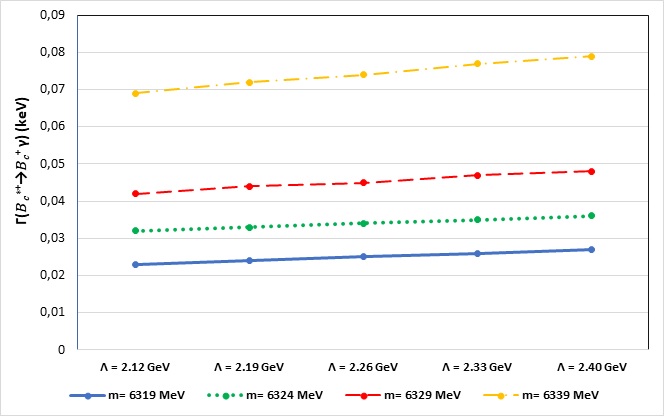}
\end{center}
\caption{
The width $\Gamma (B^{*+}_c \to B^{+}_c \gamma)$ in dependence on the choice of the $B^{*}_c$ meson mass and the size parameter $\Lambda_{B^{*}_c}$.} 
\label{fig2}
\end{figure}

The width of $\Gamma (B^{*+}_c \to B^{+}_c \gamma)$ decay strongly depends on the choice of $B^{*}_c$ meson's mass  than on the choice of $\Lambda_{B^{*}_c}$ in our calculations as expected, and shown on the Figure~\ref{fig2}. While $f_{B^{*}_c}$ leptonic decay constant depends on the choice of $\Lambda_{B^{*}_c}$.
We compared the results of widths of radiative decays of  $B^*_{q}$ mesons within the covariant confined quark model with those from other theoretical predictions in Table \ref{tab:Radiative_decay}. For $\Gamma (B^{*+}_c \to B^{+}_c \gamma)$ we used central values of assumed parameters($m_{B_{c}^{*}}=6329$ MeV  and $\Lambda_{B_{c}^{*}}=2.26$ GeV).
\begin{table}[htp] 
\caption{\label{tab:Radiative_decay}
The widths of radiative decays of  $B^*_{q}$ mesons in units of keV.}
\begin{center}
\begin{tabular}{|c|c|c|c|c|}
\hline
&$\Gamma (B^{*0} \to B^{0} \gamma)$  &$\Gamma (B^{*+} \to B^{+} \gamma)$  &$\Gamma (B^{*0}_s \to B^{0}_s \gamma)$  &$\Gamma (B^{*+}_c \to B^{+}_c \gamma)$\\
\hline
This work& $0.117\pm 0.022.$ &$0.362\pm 0.072 $&$0.094\pm 0.018$ & $0.045\pm 0.009 $\\
\cite{Ebert:2002xz,Ebert:2002pp}  &0.070     &0.19    &0.054    &0.033\\
\cite{Simonis:2018rld}&0.165     &0.520    &0.115    &0.039\\
\cite{Jena:2002is}  &0.14     &0.52    &0.06    &0.030\\
\cite{Chang:2020xvu} &$0.116\pm 0.006$  &$0.349\pm 0.018$ &$0.084^{+11}_{-9}$ &$0.049^{+28}_{-21}$\\
\cite{Priyadarsini:2016tiu,Patnaik:2017cbl}  &0.181     &0.577    &0.119    &0.023\\
\cite{Lahde:1999ih,Lahde:2002wj}  &0.0096     &0.0674    &0.148    &0.034\\
\cite{Choi:2007se,Choi:2009ai}  &0.13     &0.4    &0.068    &0.022\\
\cite{Eichten:1994gt}  &     &    &    &0.135\\
\cite{Kiselev:1994rc}  &     &    &    &0.060\\
\cite{Fulcher:1998ka}  &     &    &    &0.059\\
\cite{Nobes:2000pm}    &     &    &    &0.050\\
\cite{Monteiro:2016rzi}  &     &    &    &0.019\\
\cite{AbdElHady:2005bv}  &     &    &    &0.019\\
\hline
\end{tabular}
\end{center}
\end{table}

\section{CONCLUSION}
In this work we made naive assumptions for the $B_{c}^{*}$ meson mass and size parameter $\Lambda_{B_{c}^{*}}$ as $m_{B_{c}^{*}}=6329\pm 10$ MeV and $\Lambda_{B_{c}^{*}}=2.26\pm 0.14$ GeV . Further, using this numbers We calculated leptonic decay constants for the $B_{c}^{*}$ meson, and widths of radiative decays of  $B^*_{q}$ mesons, where $q=u/d,s,c$. In Table~\ref{tab:Bvc} and Fig.~\ref{fig2} were shown that the width $\Gamma (B^{*+}_c \to B^{+}_c \gamma)$ very sensitive to the mass $m_{B_{c}^{*}}$ as expected, and less to the size parameter $\Lambda_{B_{c}^{*}}$. While the $f_{B^{*}_c}$ leptonic decay constant strongly depends on the choice of $\Lambda_{B^{*}_c}$. There is a significant scatter in the values for the decay widths in Table~\ref{tab:Radiative_decay}.
Therefore, their experimental measurement will significantly correct the framework of the existing theoretical approaches to the description of these processes.

\section{ACKNOWLEDGEMENTS}
We would like to thank Prof. Mikhail A. Ivanov for useful discussions of some aspects
of this work.
This research has been funded by the Science Committee of the Ministry of Education and Science of the Republic of Kazakhstan (Grant No. AP09057862).

\end{document}